# Ising formulations for two-dimensional cutting stock problem with setup cost

Hiroshi ARAI
Graduate school of Science and Engineering
Ibaraki University
h.rough2019@gmail.com

Harumi HARAGUCHI
Graduate school of Science and Engineering
Ibaraki University
harumi.haraguchi.ie@vc.ibaraki.ac.jp

**Abstract**

We proposed the method that translates the 2-D CSP for minimizing the number of cuts to the Ising model. After that, we conducted computer experiments of the proposed model using the benchmark problem. From the above, the following results are obtained. (1) The proposed Ising model adequately represents the target problem. (2) Acceptance rates were low as 0.2% to 9.8% and from 21.8% to 49.4%. (3) Error rates from optimal solution were broad as 0% to 25.9%. As the future work, (1) Improve the Hamiltonian for Constraints. (2) Improve the proposed model to adjust more complex 2-D CSP and reduce the number of spins when it deals with large materials and components. (3) Conduct experiments using a quantum annealer.

## 1. Introduction

In several industries such as paper, steel, and textile, items of specified sizes are cut out from the raw material placed on a flat surface. Problems aimed at minimizing the cost of such cutting work are referred to as two-dimensional cutting stock problems (CSPs) and two-dimensional bin packing problems (BPP), on which several studies have been carried out so far.

In particular, when positioning the items to be cut from raw material of fixed width $W$ and variable height $H$, if the purpose of the problem is to minimize the height $H$, it is formulated as the strip packing problem (SPP), and various heuristic solutions have been proposed. Typical examples include the first-fit decreasing height (FFDH) method [1], in which the items are arranged in descending order of height; best-fit decreasing height (BFDH) method [2]; floor–ceiling no rotation (FCNR) method [3], in which arrangement rules are added to the FFDH method; and next-fit decreasing height (NFDH) method [1], which satisfies the guillotine cut constraint. However, several of these previous studies are aimed at maximizing the yield of raw materials as the cost target.

However, it is not just the material cost that must be considered [4]. For example, at a manufacturing site where special sheet-like materials are cut, a setup change cost is incurred for each cutting, which significantly impacts the total cost. Hence, a technique that can facilitate a reduction of the setup change cost is desired. In this paper, we propose an allocation method aimed at minimizing the number of setup changes by applying conventional methods such as the FFDH and binary tree methods for manufacturing sites that cut special sheet-like material. This method has helped achieve a reduction in the number of setup changes while ensuring a high yield [5].

Because a two-dimensional CSP is considered $\mathcal{NP}$-hard, combinatorial explosion occurs when the complexity of the problem and, in turn, the scale of the solutions increases, leading to an exponential increase in the calculation time. In contrast, the actual problem represented by the two-dimensional CSP has a short lead time and requires a high-speed solution.

Quantum computers are next-generation computers that use quantum mechanics and are implemented through two types of methods: the quantum annealing and quantum gate methods. The quantum annealer is considered to be suitable for solving combinatorial optimization problems. Therefore, various combinatorial optimization problems have been studied using quantum annealing. In a previous study on the BPP, Terada *et al.* (2018) performed Ising modeling using sequence pairs for maximizing the yield of the rectangular packing problem [6]. Further, Kida *et al.* (2000) performed Ising modeling with a formulation based on the NFDH method for the glass processing industry; they compared the performances of the quantum annealer and the classical bifurcation machine [7]. However, Ising models aimed at reducing setup change cost (= number of cuts) in the two-dimensional CSP have not yet been created.

Therefore, in this study, we propose an Ising model for the two-dimensional CSP that minimizes the number of cuts,

perform simulation experiments on benchmark problems, and compare the results with the optimal solution determined using a solver.

## 2. Ising Model

When solving a combinatorial optimization problem using a quantum annealer, it is necessary to transform it into an equation referred to as the Ising model. The Ising model is a statistical mechanics model that represents the properties of a magnetic material, and the state of the magnetic material is evaluated by calculating the upward or downward state of the spin. Spin has two forces: a local field that acts only on that spin and an interaction that acts on the connected spins. The state of the spin is updated by these forces so that the total energy is stabilized.

When the entire system is in the most stable state, the Hamiltonian $\mathcal{H}$, which represents the total energy of the Ising model, is the minimum (ground state). Equation (1) expresses this Hamiltonian. Here, it should be noted that $s_i$ and $s_j$ are the directions ($\pm 1$) of the spins $i$ and $j$, respectively, $J_{ij}$ represents the interaction, $h_i$ is the local field of spin $i$, $E$ is the number of edges, and $V$ is the number of spins:

$$\mathcal{H} = \sum_{(i,j)\in E} J_{ij} s_i s_j - \sum_{i\in V} h_i s_i \qquad (1)$$

## 3. Target Overview

We consider a two-dimensional CSP as the target problem. However, if applied as it is, the number of spins in the quantum annealer may be insufficient, as there are several constraints. Hence, some simplification is required. Fig. 1 presents an example of the allocation and specific conditions.

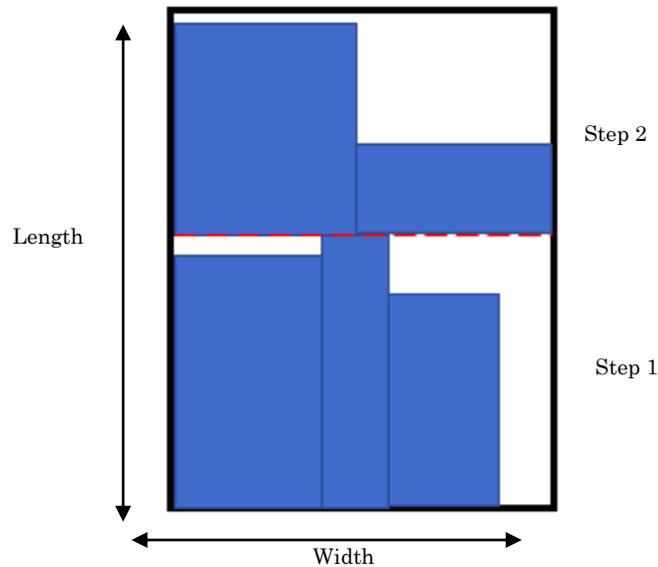

Fig. 1: Example of allocation

The main condition of this model is as follows,

1. There are infinite base materials with fixed widths and lengths.
2. The items shall be allocated from the lower left of the base material.
3. The allocated items must not protrude from the base material.
4. The allocated items must not overlap.
5. All the items are allocated from one of the base materials only once.
6. The items must not be rotated.
7. Allocation in the vertical direction of the base material is known as a step, and the length of each step is the length of the longest piece in that step.

8. Items shall not be allocated vertically in each step.
9. If there are two or more items of the same length in each step, their horizontal cutting is done all at once.
10. If the allocated piece in each base material touches the upper side of the base material, horizontal cutting is omitted.
11. If the total width of the items allocated in each step matches the width of the base material, the vertical cutting of the rightmost piece is omitted.

## 4. Ising Modeling

Table 1 lists the variables used in this paper.

Table 1: List of variables

| Symbol | Description |
|---|---|
| $Bin\_H$ | Length of base material |
| $Bin\_W$ | Width of base material |
| $I$ | Number of base materials |
| $i$ | Base material number ($i = 1, \ldots, I$) |
| $j$ | Number of vertical steps in base material ($j = 1, \ldots, Bin\_H$) |
| $K$ | Number of items |
| $k$ | Piece number ($k = 1, \ldots, K$) |
| $l$ | Spin number representing total width of items ($l = 0, \ldots, Bin\_W$) |
| $m$ | Spin number representing longest piece length ($m = 0, \ldots, Bin\_H$) |
| $n$ | Spin number representing type of piece length ($n = 1, \ldots, Bin\_H$) |
| $p$ | Spin number representing total length of longest items ($p = 0, \ldots, Bin\_H$) |
| $h_k$ | Length of piece $k$ |
| $w_k$ | Width of piece $k$ |
| $\mathcal{H}_{Hcut}$ | Hamiltonian representing number of horizontal cuts in base material |
| $\mathcal{H}_{Wcut}$ | Hamiltonian representing number of vertical cuts in base material |
| $\mathcal{H}_A$ | Hamiltonian for piece allocation |
| $\mathcal{H}_H$ | Hamiltonian representing vertical allocation constraint of a piece |
| $\mathcal{H}_W$ | Hamiltonian representing horizontal allocation constraint of a piece |
| $\mathcal{H}_{Htype}$ | Hamiltonian to obtain piece length type at each step |
| $\mathcal{H}_{Hlongest}$ | Hamiltonian to ensure that the longest piece in each step is indeed the longest |
| $\sigma$ | Objective function parameters |
| $\lambda_A$ | $\mathcal{H}_A$ parameters |
| $\lambda_H, \mu_H$ | $\mathcal{H}_H$ parameters |
| $\lambda_W, \mu_W$ | $\mathcal{H}_W$ parameters |
| $\sigma_t$ | $\mathcal{H}_{Htype}$ parameters |
| $\sigma_l, \lambda_l$ | $\mathcal{H}_{Hlongest}$ parameters |

Further, there are five types of spins as follows:

$$s_{i,j,k} = \begin{cases} 1 & \text{Piece } k \text{ is allocated to } j\text{th step of base material } i \\ 0 & \text{Otherwise} \end{cases}$$

$$sw_{i,j,l} = \begin{cases} 1 & \text{Total width of items allocated to } j\text{th step of base material } i \text{ is } l \\ 0 & \text{Otherwise} \end{cases}$$

$$sh_{i,j,m} = \begin{cases} 1 & \text{Longest piece in } j\text{th step of base material } i \text{ is m} \\ 0 & \text{Otherwise} \end{cases}$$

$$shc_{i,j,n} = \begin{cases} 1 & \text{Piece of length } n \text{ is allocated in } j\text{th step of base material } i \\ 0 & \text{Otherwise} \end{cases}$$

$$sht_{i,p} = \begin{cases} 1 & \text{Total length of longest items of each step of base material } i \text{ is } p \\ 0 & \text{Otherwise} \end{cases}$$

The full Hamiltonian of the proposed Ising model is expressed as follows:

$$\mathcal{H} = \mathcal{H}_{Hcut} + \mathcal{H}_{Wcut} + \mathcal{H}_A + \mathcal{H}_H + \mathcal{H}_W + \mathcal{H}_{Hlongest} + \mathcal{H}_{Htype} \quad (2)$$

Where $\mathcal{H}_{Hcut}$ and $\mathcal{H}_{Wcut}$ are Hamiltonians that represent objective functions and $\mathcal{H}_A, \mathcal{H}_H, \mathcal{H}_W, \mathcal{H}_{Hlongest},$ and $\mathcal{H}_{Htype}$ are Hamiltonians that represent constraints. $\mathcal{H}_{Hcut}$ and $\mathcal{H}_{Wcut}$ are represented as follows:

$$\mathcal{H}_{Hcut} = \sigma \sum_{i=1}^{I} \left( \sum_{j,n=1}^{Bin\_H} shc_{i,j,n} - sht_{i,Bin_H}(1 - sht_{i,0}) \right) \quad (3)$$

$$\mathcal{H}_{Wcut} = \sigma \sum_{i=1}^{I} \sum_{j=1}^{Bin\_H} \left( \sum_{k=1}^{K} s_{i,j,k} - sw_{i,j,Bin_W}(1 - sw_{i,j,0}) \right) \quad (4)$$

Equation (3) represents the number of cuts in the horizontal direction and satisfies Constraints 9 and 10 listed in Section 3. For each piece length type at each step, the number of cuts is incremented, and when the sum of lengths of the longest items at each step matches the length of the base material, the number of cuts is omitted once.

Equation (4) expresses the number of cuts in the vertical direction and satisfies Constraint 11 listed in Section 3. The number of cuts is incremented for each piece at each step, and when the total width of the items allocated to each step matches the width of the base material, the number of cuts is omitted once.

$\mathcal{H}_A$ is expressed as follows:

$$\mathcal{H}_A = \lambda_A \left( \sum_{i=1}^{I} \sum_{j=1}^{Bin\_H} s_{i,j,k} - 1 \right)^2 \quad \forall k \quad (5)$$

Equation (5) expresses a constraint on the number of times items allocated and satisfies Constraint 5 listed in Section 3. Its value is the lowest when all the items are allocated to one of the base materials only once.

$\mathcal{H}_H$ and $\mathcal{H}_W$ are expressed as follows:

$$\mathcal{H}_H = \lambda_H \left( 1 - \sum_{p=0}^{Bin\_H} sht_{i,p} \right)^2 + \mu_H \left( \sum_{p=0}^{Bin\_H} p \cdot sht_{i,p} - \sum_{j=1,m=0}^{Bin\_H} m \cdot sh_{i,j,m} \right)^2 \quad \forall i \quad (6)$$

$$\mathcal{H}_W = \lambda_W \left( 1 - \sum_{l=0}^{Bin\_H} sw_{i,j,l} \right)^2 + \mu_W \left( \sum_{l=0}^{Bin\_W} l \cdot sw_{i,j,l} - \sum_{k=1}^{K} w_k \cdot s_{i,j,k} \right)^2 \quad \forall i,j \quad (7)$$

Equation (6) represents the constraint for preventing the items from protruding vertically from the base material, and its value is the lowest when there is only one spin with $sht = 1$ in each base material and the subscript $p$ of the spin with $sht = 1$ matches the sum of the subscripts $m$ of the $sh$ at each step.

Equation (7) expresses the constraint for preventing the items from protruding horizontally from the base material, and its value is the lowest when there is only one spin with $sw = 1$ at each step for each base material and the subscript $l$ of the spin with $sw = 1$ matches the total width of the items allocated to the target step.

Equations (6) and (7) satisfy Condition 3 listed in Section 3. $\mathcal{H}_{Htype}$ is expressed as follows:

$$\mathcal{H}_{Htype} = -\sigma_t \sum_{n=1}^{Bin\_H} (1 - shc_{i,j,n}) \left(1 - 2 \sum_{k|h_k=n} s_{i,j,k}\right) \forall i,j \quad (8)$$

Although Equations (3)–(7) can be used to represent a two-dimensional CSP that minimizes the number of cuts, the spin $shc$ cannot always correctly represent the piece length type at each step. Therefore, a constraint is set on $shc$ by Equation (8). This equation yields the minimum value when the values of all the variables $s_{i,j,k}$ with $h_k = n$ become 0 at each step for each base material, leading to $shc = 0$, and when one of these values becomes 1, leading to $shc = 1$.

$\mathcal{H}_{Hlongest}$ is expressed as follows:

$$\mathcal{H}_{Hlongest} = \lambda_l \left(1 - \sum_{m=0}^{Bin\_H} sh_{i,j,m}\right)^2 - \sigma_l \sum_{n=1}^{Bin\_H} n \cdot shc_{i,j,n} sh_{i,j,m|m=n} \quad \forall i,j \quad (9)$$

Similar to $shc$, the spin $sh$ cannot always correctly represent the longest piece at each step only through Equations (3)–(8). Therefore, a constraint is set on $sh$ by Equation (9). This equation yields the minimum value only when the $sh$ with the subscript $m$ that matches the longest of the items allocated at each step for each base material becomes 1.

## 5. Computer experiment

- Experimental conditions

To confirm the validity of the proposed Ising model, we performed a computer experiment. Because the target problem is a benchmark problem [8], it is considered as an instance that satisfies the conditions specified in Table 2.

However, for the sake of simplicity, the length of the base material $Bin\_H$ is considered as a variable, and it is assumed that a cut is always made in line with the longest piece among the allocated items. Thus, the full Ising model of this experiment is represented as follows:

$$\mathcal{H} = \mathcal{H}_{Hcut} + \mathcal{H}_{Wcut} + \mathcal{H}_A + \mathcal{H}_W + \mathcal{H}_{Htype} \quad (10)$$

Table 2: Experimental conditions for the benchmark problem

| Name | Value |
| --- | --- |
| $Bin\_H$ | 10 |
| $Bin\_W$ | 10 |
| $K$ | 20 |

For the experiment, Equation (10) is converted into a QUBO matrix, which is known to have a one-to-one correspondence with the Ising model. The QUBO matrix is applied to QBsolv, which can perform optimization calculations through tabu search, and neal, which can perform optimization calculation using simulated annealing [9]. The number of trials was set to 1000. Table 3 lists each parameter that was set by the preliminary experiment.

Table 3: Parameter settings

| Variable identifier | Value |
| --- | --- |
| $\sigma$ | 1000 |
| $\sigma_t$ | 5000 |
| $\lambda_A$ | 500000 |
| $\lambda_W$ | 500000 |
| $\mu_W$ | 10000 |
| Number of trials | 1000 |

- Experimental results

We now consider the acceptance rate and accuracy of the experimental results. Further, we calculate the optimal solution of the target benchmark problem using Google OR-tools and compare the two sets of results.

Tables 4 and 5 indicate that the error rates of the best solution calculated using the proposed Ising model and the optimal solution calculated using Google OR-tools were 0%–25.9%, respectively, obtained using both QBsolv and neal. The optimal solutions were obtained in Instances 9 and 3 in QBsolv and neal, respectively. The acceptance rates were 0.2%–9.8% and 21.8%–49.4%, respectively. In particular, several solutions that violated the constraints were calculated using QBsolv. These solutions can be improved by adjusting the parameters or modifying the Ising model. Although the parameters $\mathcal{H}_{Htype}$ and $\mathcal{H}_{Hlongest}$ of the Ising model are constraints, their minimum values change depending on the problem and others parameters, and they are effectively objective functions that impose constraints on other parameter settings. If a model could be constructed such that the values of these Hamiltonians could reach a minimum of 0, the range of acceptable values for the other parameters would become wider, and flexible response would become possible.

Further, the reason for the difference in acceptance rate, depending on the instance, is that if the number of items having widths exceeding 50% of the width of the base material (number of high-width items) is high, the ratio of the solution space to the search space becomes small, and the probability of obtaining an acceptable solution within the execution time becomes low. Figs. 2 and 3 present the distribution maps of the acceptance rates and the numbers of high-width items at each instance obtained using QBsolv and neal, respectively. The broken lines in the figures represent the linear approximation curves. These figures indicate that the acceptance rate of the solution decreases proportionally with the number of high-width items.

Table 4: Comparison of the experimental results obtained using QBsolv and the optimal solution

| No. | Best value | Acc. rate | Av. | Var. | SD | Opt. |
| --- | --- | --- | --- | --- | --- | --- |
| 1 | 35 | 3.2 | 37.688 | 1.152 | 1.073 | 32 |
| 2 | 34 | 5.7 | 37.702 | 1.613 | 1.270 | 29 |
| 3 | 35 | 0.7 | 37.000 | 1.143 | 1.069 | 34 |
| 4 | 34 | 3.1 | 37.839 | 1.361 | 1.167 | 27 |
| 5 | 34 | 0.9 | 34.889 | 0.321 | 0.567 | 31 |
| 6 | 34 | 0.2 | 35.000 | 1.000 | 1.000 | 31 |
| 7 | 34 | 2.7 | 37.370 | 1.048 | 1.024 | 27 |
| 8 | 35 | 9.8 | 38.622 | 1.357 | 1.165 | 28 |
| 9 | **32** | 1.0 | 36.900 | 3.090 | 1.758 | **32** |
| 10 | 35 | 0.9 | 36.778 | 1.284 | 1.133 | 28 |

Table 5: Comparison of the experimental results obtained using neal and the optimal solution

| No. | Best value | Acc. rate | Av. | Var. | SD | Opt. |
| --- | --- | --- | --- | --- | --- | --- |
| 1 | 34 | 36.3 | 37.788 | 1.164 | 1.079 | 32 |
| 2 | 34 | 42.4 | 37.807 | 1.081 | 1.039 | 29 |
| 3 | **34** | 26.2 | 37.206 | 0.706 | 0.840 | **34** |
| 4 | 34 | 40.0 | 37.800 | 1.040 | 1.020 | 27 |
| 5 | 32 | 27.3 | 35.150 | 0.824 | 0.908 | 31 |
| 6 | 33 | 21.8 | 35.459 | 0.459 | 0.678 | 31 |
| 7 | 34 | 46.3 | 37.659 | 1.110 | 1.054 | 27 |
| 8 | 34 | 49.4 | 38.690 | 1.340 | 1.160 | 28 |
| 9 | 34 | 30.5 | 37.230 | 0.774 | 0.880 | 32 |
| 10 | 32 | 26.5 | 36.758 | 1.157 | 1.076 | 28 |

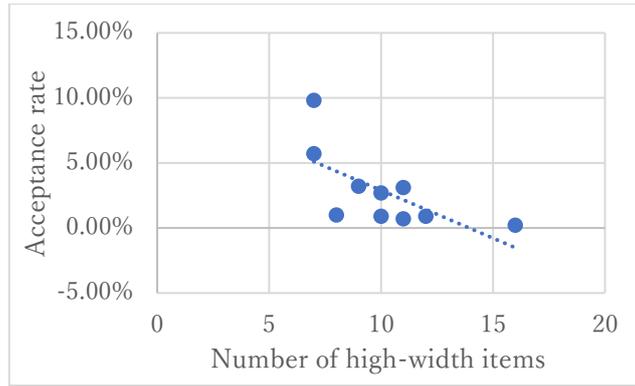

Fig. 2: Distribution of the acceptance rates obtained using QBsolv at each instance and the numbers of high-width items at those instances

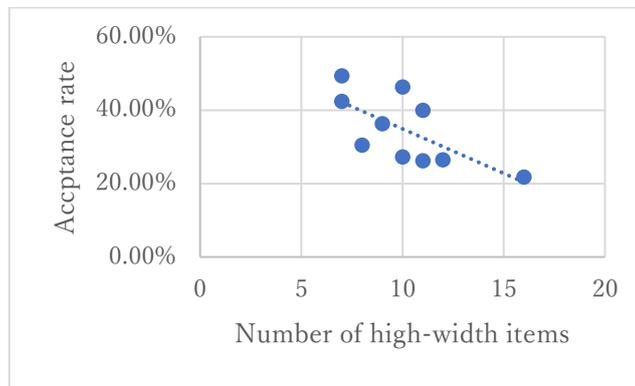

Fig. 3: Distribution of the acceptance rates obtained using neal at each instance and the numbers of high-width items at those instances

## 6. Conclusion

In this paper, we proposed an Ising model for the purpose of minimizing the number of cuts in a two-dimensional CSP. The proposed Ising model worked appropriately, and as the number of cuts decreased, the energy consumed also reduced. However, in the experiment, the acceptance rates were 0.2%–9.8% for QBsolv and 21.8%–49.4% for neal, and the error rate of the optimal solution was 0%–25.9%, indicating a wide range. Further, in the proposed method, the constraints for the allocation of items in the vertical and horizontal directions were realized by preparing spins based on the size of the base material. However, as the size of the base material increases, the number of spins to be secured also increases, making it difficult to apply to the quantum annealer. The following can be listed as future challenges:

- Improving the constraints used in the Ising model
- Adjusting the parameters further
- Considering the omission of vertical cutting
- Devising an Ising model that can be applied to situations that use large-sized base materials
- Application to the quantum annealer

## Acknowledgements

We would like to express our gratitude to Mr Akira Miki of DENSO Corporation for sharing his valuable opinions on this concept with us.